\documentclass[aps,prl,amssymb,nofootinbib,twocolumn]{revtex4}
\usepackage{amsfonts,amsmath,amsbsy,color}
\usepackage{textcomp}
\usepackage[]{graphics,graphicx}
\usepackage[version=3]{mhchem}
\usepackage{subfigure}
\begin{document}
\title[Holomorphic HF]{Holomorphic Hartree--Fock theory: an inherently multireference approach}
\author{Hugh~G.~A.~Burton}
\author{Alex~J.~W.~Thom}
\email{ajwt3@cam.ac.uk}
\affiliation{University Chemical Laboratory, Lensfield Road, Cambridge CB2 1EW, United Kingdom}
\date{\today}
\begin{abstract}
We investigate the existence of holomorphic Hartree-Fock solutions using a revised SCF algorithm.  
We use this algorithm to study the Hartree-Fock solutions for $\ce{H2}$ and $\ce{H4^2+}$ and report the emergence of holomorphic solutions at points of symmetry breaking.  
Finally, we find these holomorphic solutions for $\ce{H4}$ and use them as a basis for Non-Orthogonal Configuration Interaction at a range of rectangular geometries and show them to produce energies in good agreement with Full Configuration Interaction.

\end{abstract}
\maketitle
\section{Introduction}
The Hartree-Fock (HF) approximation provides the bedrock of modern wavefunction based methods in Quantum Chemistry.
Representing the wavefunction as a single Slater Determinant constructed from the product of one electron spin orbitals, a Self-Consistent Field (SCF) method is used to minimise the energy with respect to variation in the basis orbitals, thus solving the non-linear Roothaan-Hall equations\cite{Roothaan,Hall}. 
It is a mean-field method whereby the electron-electron repulsion experienced by one electron is the average of the repulsion from all other electrons, and hence its solutions do not describe electron correlation effects.
This is readily observed when studying the Restricted HF solution to the ground state of $\ce{H2}$ which does not produce the correct behaviour at dissociation.
On the simplest level such behaviour at dissociation may be corrected by using an unrestricted Hartree-Fock (UHF) method, thus allowing the spatial parts of the $\alpha$ orbitals to differ to that of the $\beta$ orbitals.
Beyond this, the solutions of the Hartree-Fock method may be used as a reference for further correlation methods.

Configuration Interaction (CI) methods provide the most powerful solution to the problem of electron correlation.
These take a linear combination of excited state determinants --- obtained by replacing occupied orbitals with virtual orbitals --- to introduce a term depending on the inter-electron distance into the wavefunction. 
The Full CI method uses all possible replacement determinants and yields the exact solution to the Hamiltonian in the Hilbert Space of the system.
Inevitably this method is very computationally expensive, scaling factorially with the number of electrons N, \cite{FCIScaling} and so the use of a limited number of excited determinants is common in methods such as CISD $\cal{O}$($N^6$) which uses just single and double excitations. 

Recently, one of us has studied the existence of multiple solutions to the SCF equations using a method known as Metadynamics \cite{Metadynamics}. 
These solutions have been investigated by a number of other authors \cite{BesleyGill_09JCP, GillMOM, ExcitedNO, ExcitedDFT} and it is thought that they may provide a good representation of the excited states of molecules.
These low-energy SCF solutions have already proved applicable for a Restricted Active Space Self-Consistent Field method,\cite{RASSCF} and a study on $\ce{LiF}$ and $\ce{O3}$\cite{NOCIMetadynamics} found that they show similar properties to diabatic molecular states of these molecules. 
In particular they do not obey non-crossing rules and they maintain a similar electronic structure as the molecular geometry is varied.

\begin{figure}[h!]
\includegraphics[scale=0.45]{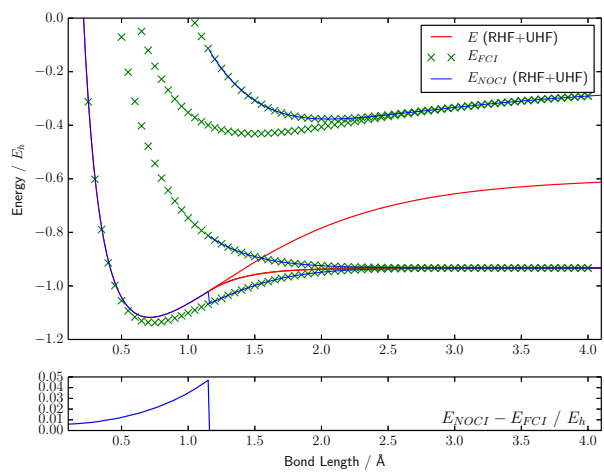}
\caption{The lowest energy RHF and UHF solutions for $\ce{H2}$ using a STO-3G basis set are plotted and the disappearance of the UHF solutions may be observed at around 1.2\r{A}. This point is known as the Coulson-Fischer point. These states are used as a basis for NOCI and a discontinuity in the ground state NOCI energy at the Coulson-Fischer point results from the disappearance of the UHF.}
\label{H2_normal}
\end{figure}
The SCF states are not in general orthogonal, but may still be used in a configuration interaction calculation, and these states provide a suitable basis for Non-Orthogonal Configuration Interaction (NOCI) calculations\footnote{If very many SCF solutions (of the order of the size of the Hilbert space) are used in NOCI then there may be linear dependencies. However, we propose to only use the low-lying SCF states, which are generally significantly fewer in number than the total size of the Hilbert space, so we not envisage this to be a problem in practice.} yielding adiabatic states which show similar properties to the states produced by more expensive CI methods\cite{NOCIMetadynamics}.
In particular, these states reproduce avoided crossings\cite{NOCIMetadynamics}, conical intersections and we believe them to maintain size-consistency because they are formed from a basis of size-consistent SCF solutions. 
By using this basis, NOCI differs from a truncated CI, whose size-inconsistency derives from the truncation at a fixed excitation level for a single reference.

For this method to be useful it is imperative that these SCF states exist across all geometries of the system, yet the coalescence and disappearance of states is widely reported for a number of molecules\cite{CoulsonFischer,CohenH4}.
Even for the modest $\ce{H2}$, it is found that the lowest energy UHF state coalesces (Figure \ref{H2_normal}) with the lowest energy RHF state at the Coulson-Fischer point and disappears at shorter bond lengths, preventing the use of these states as a basis for NOCI calculations.
Whilst the total number of solutions to the SCF equations is unknown (and one particular study by Fukutome \cite{Fukutome_71PTP} predicts it to scale dramatically with system size), to be useful the number of solutions should remain constant at all geometries.
Using a revised Holomorphic Hartree-Fock theory\cite{HiscockHoloHF} we have located holomorphic-UHF solutions (hUHF) for STO-3G $\ce{H2}$ at bond lengths shorter than the Coulson-Fischer point.
These correspond exactly to the UHF solutions where these exist, hence providing a constant number of solutions at all geometries.
We believe this to also be the case for larger molecules.

To find these hUHF solutions, one must solve a holomorphized Schr{\"o}dinger equation whereby the construction of the density matrix is adjusted by removing all complex conjugates, producing a non-Hermitian Hamiltonian. 
Previous methods using non-Hermitian Hamiltonians have been used to study metastable electronic states and Feshbach resonances \cite{McCurdy1980,McCurdy2015,NH-HF,ComplexHF,ShapeResonances} whilst complex valued orbital coefficients have been used in conjunction with a standard Hermitian matrix to explore symmetry broken RHF states.\cite{RHFcomplex}
Our proposed holomorphic method combines a non-Hermitian Hamiltonian with complex orbital coefficients which are solved to find hUHF solutions.
These solutions are a basis for NOCI which provide a good description of the system ground state.

In this paper, we study the existence and properties of these hUHF states further and propose a holomorphic SCF algorithm as a procedure for finding such solutions.
This algorithm is used to find holomorphic solutions for $\ce{H2}$ in a minimal STO-3G and 6-31G* basis set and these are then used as a basis for a NOCI calculation before being compared to the FCI solutions in the same basis set. 
Finally, we apply the method to $\ce{H4^2+}$ and $\ce{H4}$ in both a square and rectangular geometry and compare our results for $\ce{H4}$ to those of other correlated methods.

\section{Holomorphic SCF}
Under a conventional SCF algorithm, the initial step is to guess a set of trial coefficients for $N$ orbitals (denoted $i, j,\dots$ expressed in a basis of $M$ functions (denoted $\mu,\nu,\dots$).
These are used to generate a density matrix which is in turn used to form a Fock matrix.
This Fock matrix is diagonalised to produce a new set of orbital coefficients and the occupied orbitals are selected.
Using this new set of coefficients, the density matrix may be regenerated and the process repeated until self-consistency is reached.
The convergence of this process may be accelerated using Pulay's Direct Inversion of the Iterative Subspace (DIIS) \cite{PulayDIIS1,PulayDIIS2} extrapolation, which uses a linear combination of previously calculated Fock matrices to minimise a DIIS error vector (defined as $ \boldsymbol{F} \boldsymbol{P} - \boldsymbol{P} \boldsymbol{F}$).

The holomorphic UHF solutions can be found by locating the stationary points on the holomorphic energy surface.
In practice this is just an adjustment to the Hartree-Fock energy functional, achieved by removing all complex conjugates.
Defining the holomorphic density matrix as
\\$$ {\widetilde{P}}^{\mu \nu} = \sum_i^N C_i^\mu C_i^\nu,$$
we replace the normal density matrices in the conventional Hartree--Fock energy functional (see e.g.~Ref.~\onlinecite{Metadynamics} for notation) with density matrices of this form to give a holomorphic energy functional,
\begin{widetext}
\begin{align*}
\widetilde{E} = \sum_{\mu\nu}{\widetilde{P}}^{\mu\nu}h_{\mu\nu} + \frac{1}{2}\sum_{\mu\nu\sigma\tau}[{\widetilde{P}}^{\mu\nu}(\mu\nu|\sigma\tau){\widetilde{P}}^{\sigma\tau}-(\mu\sigma|\nu\tau)(^{\alpha}{\widetilde{P}}^{\mu\nu}\ ^{\alpha}{\widetilde{P}}^{\sigma\tau}+\ ^{\beta}{\widetilde{P}}^{\mu\nu}\ ^{\beta}{\widetilde{P}}^{\sigma\tau})]
\end{align*}
\end{widetext}
Throughout the holomorphic SCF procedure, we have normalised the orbitals using a method which also removes the complex conjugate, and require that for each orbital $i$, 
$$ 1 = \sum_\mu^M C^\mu_i \times C^\mu_i.$$
Removal of these complex conjugates causes the functional to become a polynomial in the holomorphic density matrix\cite{HiscockHoloHF}.
Taking inspiration from the single variable case\cite{HiscockHoloHF}, where the Fundamental Theorem of Algrebra guarantees that there exists a constant number of solutions at all geometries, we have investigated whether the number of solutions also remains constant in this holomorphic generalisation of Hartree--Fock theory.

When using the holomorphic energy functional, the holomorphic density and Fock matrices are no longer Hermitian but instead complex symmetric allowing the Fock matrix eigenvalues to become complex and thus selection of new occupied orbitals may no longer be made using the Aufbau principle. 
Instead we use the Maximum Overlap Method developed by Gill \textit{et al.}\cite{GillMOM}, selecting the new orbitals as the ones with the greatest projection into the space spanned by the previous set.
This also makes it much easier to follow excited state solutions across geometries.
Being complex symmetric matrices, the left and right eigenvectors are the same so we arbitrarily choose to use the right eigenvectors\cite{eigenvectors}.

\begin{figure*}
\centering
\includegraphics[scale=0.42]{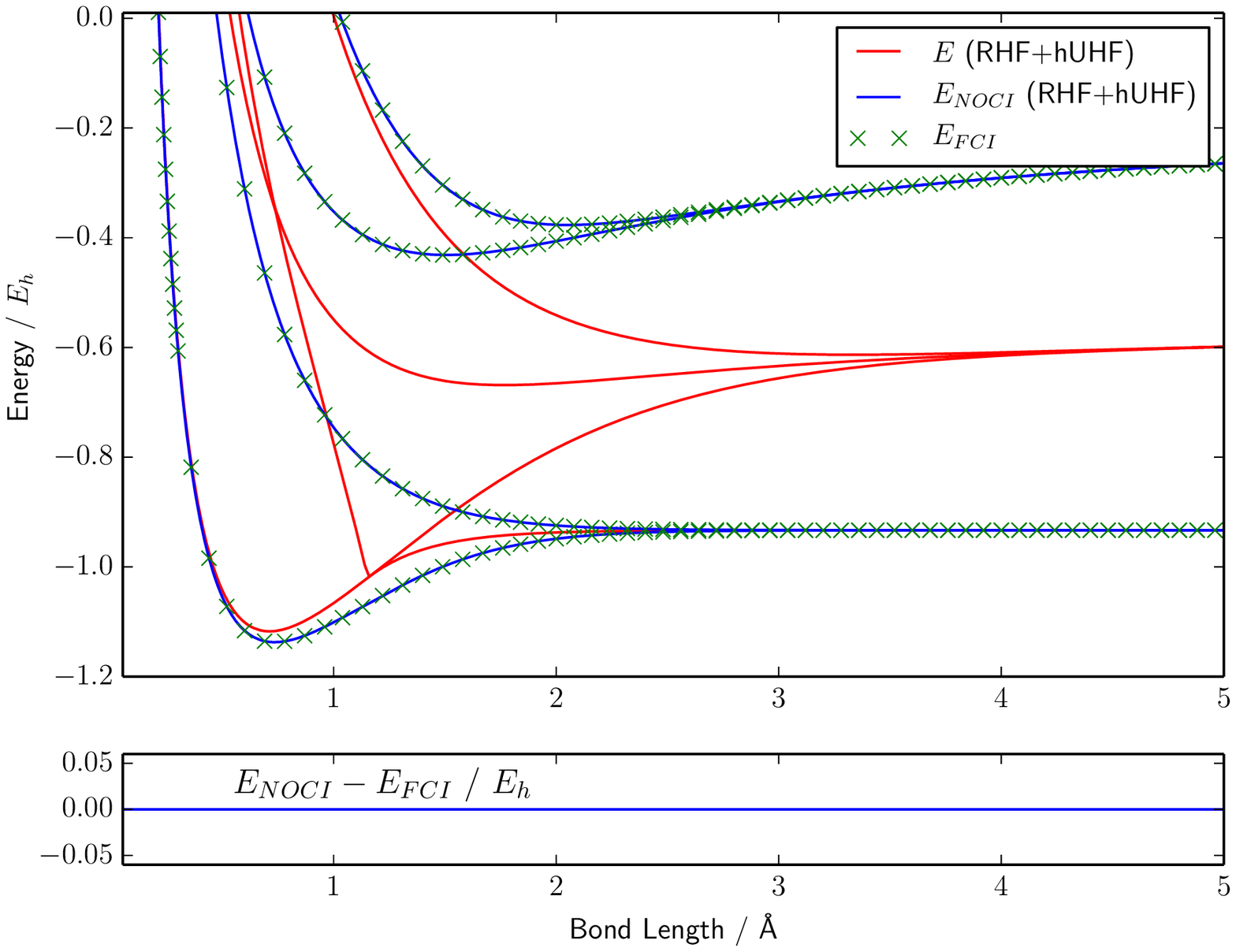}
\includegraphics[scale=0.42]{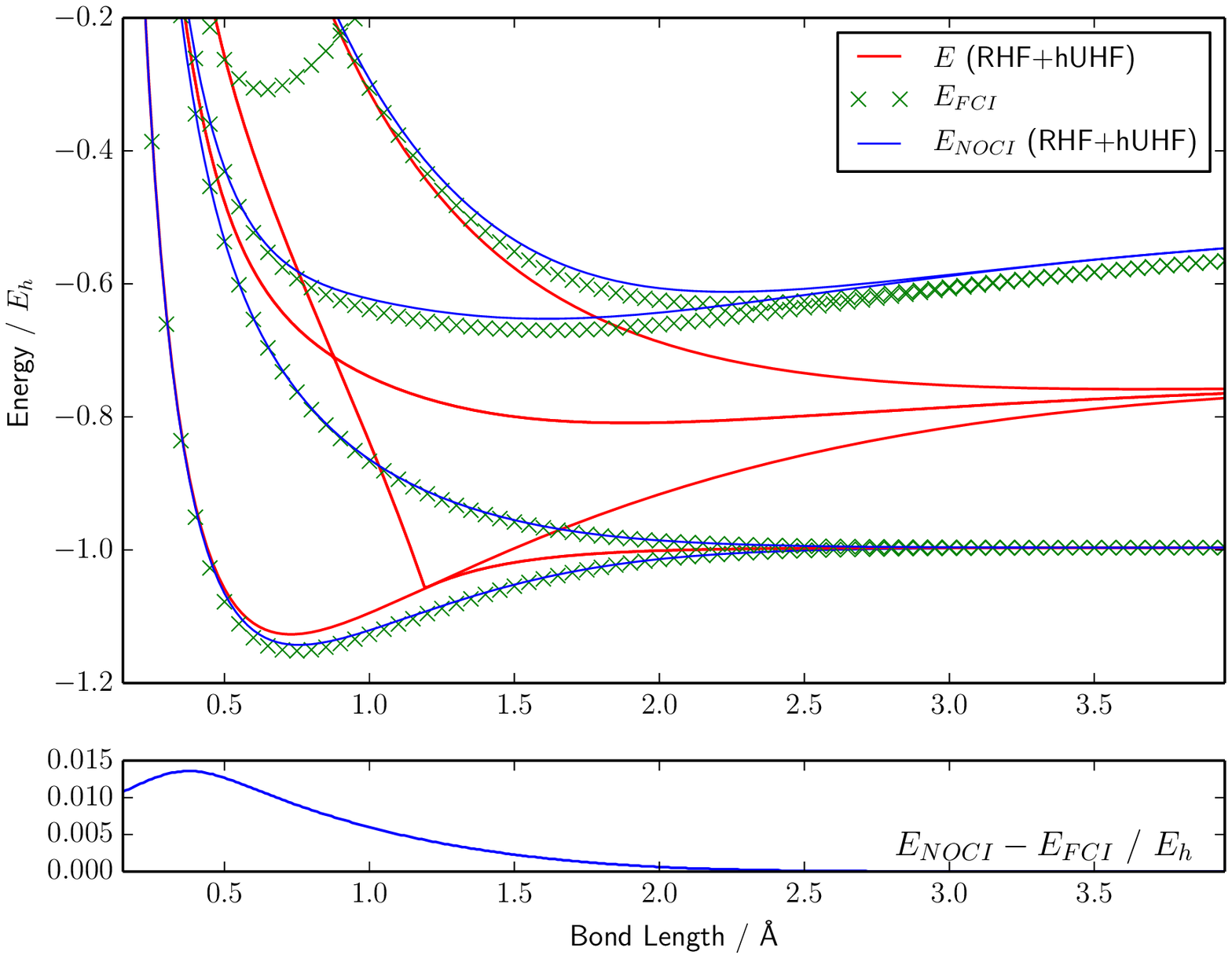}
\caption{The hUHF, NOCI and FCI states are plotted for $\ce{H2}$ in a STO-3G (left) and 6-31G* (right) basis. The holomorphic solutions can be rising out of the Coulson-Fischer point where the normal UHF solutions disappear. For the STO-3G basis, the NOCI states correspond to the FCI states since their basis spans the full Hilbert space. For 6-31G*, there is still a very good correspondence and the observed NOCI states are smooth.}
\label{fig-H2}
\end{figure*}

In summary, our proposed holomorphic SCF algorithm proceeds as follows:
\begin{enumerate}
\item Begin with a (possibly complex) guess for coefficients $C^\mu_{i}$.
\item Form the one-particle holomorphic density matrix ${\widetilde{P}}^{\mu\nu}=\sum_i^N C^\mu_{i} C^{\nu}_i$.
\item Generate the $\alpha$ and $\beta$ Fock matrices  

${{^{\alpha}\widetilde{F}}_{\mu\nu}={h}_{\mu\nu}+\sum_{\sigma\tau}^N [{{\widetilde{P}}^{\sigma\tau}}(\mu\nu|\sigma\tau)-{^{\alpha}\widetilde{P}}_{\sigma\tau}(\mu\sigma|\nu\tau)]}$.

${{^{\beta}\widetilde{F}}_{\mu\nu}={h}_{\mu\nu}+\sum_{\sigma\tau}^N [{{\widetilde{P}}^{\sigma\tau}}(\mu\nu|\sigma\tau)-{^{\beta}\widetilde{P}}_{\sigma\tau}(\mu\sigma|\nu\tau)]}$.
\item Generate new orbitals from the right eigenvectors of the Fock matrix.
\item Form the holomorphized overlap matrix of old and new orbitals. 

$ {\widetilde{O}_{ij}} = {\sum_\mu^M}\ ^\textrm{old}{C}^\mu_i\ ^\textrm{new}{C}^\mu_j$.

This overlap matrix is used to select the new orbitals through the Maximum Overlap Method.
\item Repeat until self-consistency is reached.
\end{enumerate}
We try this algorithm with a large number of different initial guess sampled randomly to find as many stationary points as possible.
Once convergence has been reached, the true energy of the holomorphic solutions may be found by orthonormalizing the orbitals and then using the normal energy functional; these states may then be used alongside standard RHF and UHF solutions as a basis for configuration interaction methods.
This step appears to be a principal difference to the non-Hermitian SCF methods proposed by McCurdy and Head-Gordon\cite{McCurdy2015,NH-HF} whereby the ``true'' exterior scaled wavefunction is evaluated using a transformed variable.
Since the known RHF and normal UHF states are also stationary points on the holomorphic potential energy surface, they are also found by the holomorphic SCF algorithm, and included in the CI.
 
\section{Computational Details}
Throughout the computational work, the necessary integrals were generated using a modified version of Q-Chem 4.3\cite{QChem} and the SCF algorithm was implemented with additional processing using SciPy\cite{SciPy}. 
All figures were plotted with matplotlib\cite{Hunter:2007}. 

\section{Holomorphic solutions to $\ce{H2}$}
The use of excited SCF solutions as a basis for configuration interaction calculations is not possible even for $\ce{H2}$.
Breakdown of the RHF solution to yield a symmetry broken UHF solution (where the $\alpha$ and $\beta$ electrons have become isolated on separate atoms) gives the Coulson-Fischer\cite{CoulsonFischer} point and an inconsistent number of states across all geometries.
Using our holomorphic SCF algorithm we have found hUHF solutions with complex coefficients which appear to correspond to these UHF solutions beyond the Coulson-Fischer point.

Under a STO-3G basis set (corresponding to one atomic orbital centred on each H atom) we used QChem to generate the $\sigma_{g}$ and $\sigma_{u}$ symmetry orbitals which were then used as the basis set for the holomorphic SCF algorithm. 
We identified two RHF and two UHF solutions (each UHF solution is doubly degenerate) and these are shown on the left-hand graph in Figure \ref{fig-H2}, the lowest of each matches those found previously in Ref. \onlinecite{HiscockHoloHF}. 
We note the existence of a higher doubly degenerate RHF state corresponding to $\ce{H^+ \dots H^-}$ and $\ce{H^- \dots H^+}$ at dissociation however we do not search for these as accurate results may still be achieved with the states plotted in Figure \ref{fig-H2}.
At the Coulson-Fischer point, a state can be seen rising sharply out of the UHF solution: this is a pair of holomorphic solutions which have complex coefficients.
Whilst the cusp in the real energy of the hUHF state may appear distressing, we have previously found that the holomorphic energy is smooth and continuous\cite{HiscockHoloHF}.
In the Non-Orthogonal CI (NOCI) method, the six hUHF solutions are used to form a Hamiltonian matrix which is then diagonalized to produce a new set of new energy eigenvalues.
Significantly, the NOCI solutions under the STO-3G basis show a smooth ground state energy which corresponds perfectly with the lowest energy FCI state.
However, this is to be expected as under this minimal basis set the Hilbert space is completely spanned by the hUHF and RHF solutions and thus the NOCI method becomes equivalent to the FCI.

With this is mind, we repeat the calculations using the larger 6-31G* basis, corresponding to two atomic orbitals centred on each atom.
The set of six hUHF solutions are still used as a basis for NOCI but now the Hilbert space contains 16 determinants.
These results are shown on the right-hand graph of Figure \ref{fig-H2}.
Despite the hUHF solutions spanning only a subspace of the Hilbert Space, the NOCI states remain smooth as predicted\cite{HiscockHoloHF} and each one corresponds closely to one of the FCI states.
The symmetries of these states may be identified (from the bottom) as $^1\Sigma_g^+, ^3\!\Sigma_u^+, ^1\!\Sigma_u^+, ^1\!\Sigma_g^+$.

\section{Symmetry Breaking in $\ce{H4^2+}$}
\begin{figure}
\includegraphics[scale=1.2]{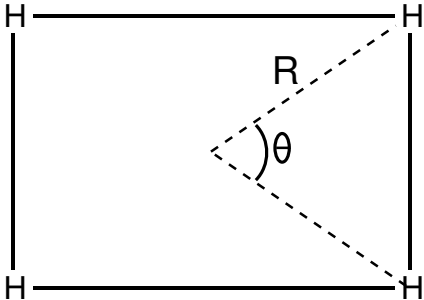}
\caption{In this model, the size of the rectangle is dictated by the parameter R whilst $\theta$ quantifies the degree of asymmetry.}
\label{H4_model}
\end{figure}
\begin{figure*}
\includegraphics[scale=0.42]{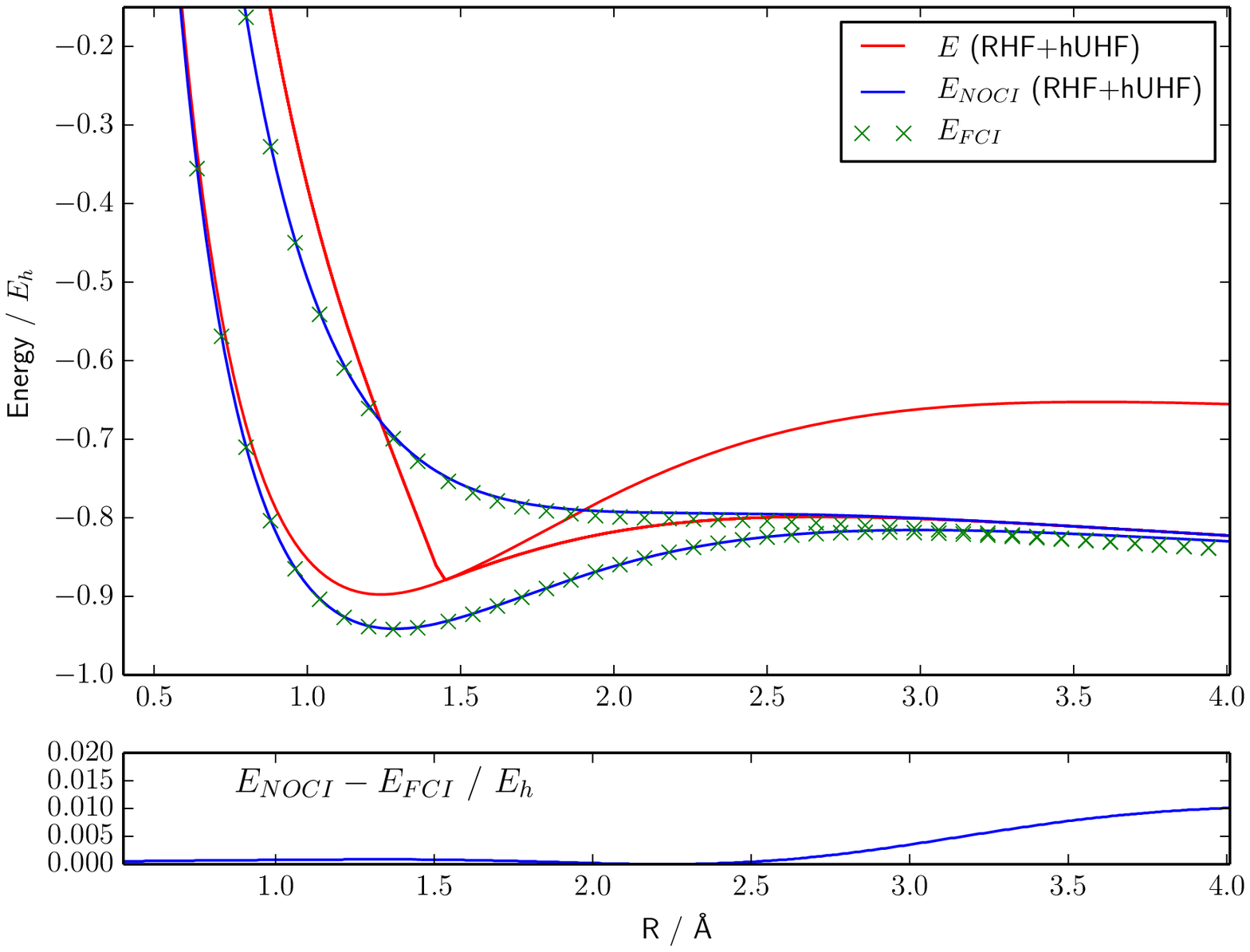}
\includegraphics[scale=0.42]{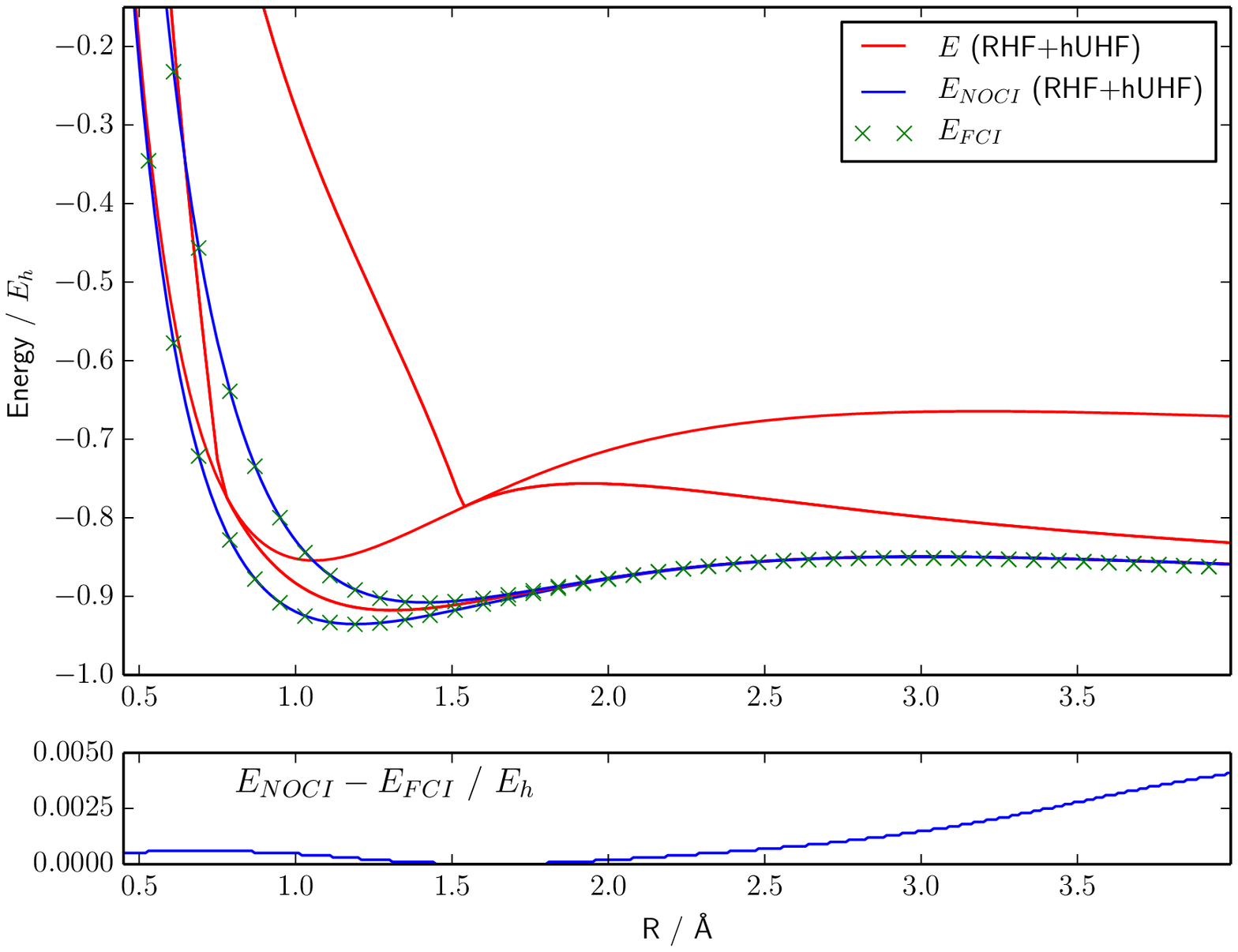}
\caption{The hUHF and NOCI states are plotted along with the ground FCI state for $\ce{H4^2+}$ using a STO-3G basis for a square (left) and rectangular (right) geometry with aspect ratio 5:7. Two hUHF states can be seen for the rectangular geometry, both doubly degenerate and corresponding to the expected UHF state beyond the Coulson-Fischer point. In the square geometry these two states become degenerate to give a single symmetry broken state with four-fold degeneracy.}
\label{H4_rect}
\end{figure*}
$\ce{H2}$ is not the only molecule which exhibits disappearing UHF solutions, in fact many diatomics such as $\ce{F2}$ also have a Coulson-Fischer point and symmetry breaking of states has been observed in yet more complex molecules.
Recently, Cohen \textit{et al.} \cite{CohenH4} have shown that symmetry broken states exist in the ground RHF state of $\ce{H4^2+}$ with a rectangular arrangement.
Using our holomorphic SCF method, we used a STO-3G basis to investigate this RHF state and its symmetry broken UHF states and these are plotted in Figure \ref{H4_rect} for a square geometry and a rectangle with aspect ratio 5:7.
Under this model it is possible to simulate the interaction of two $\ce{H2^+}$ molecules with very large ratios representing essentially non-interacting molecules and the square representing a strongly interacting system.

\begin{figure}
\includegraphics[scale=0.45]{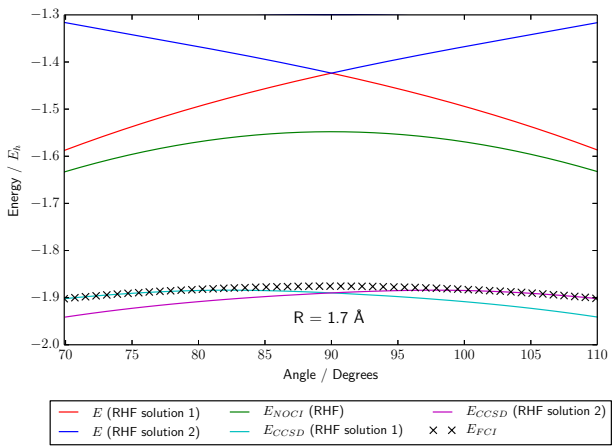}
\caption{The lowest energy RHF (red and blue) solutions for $\ce{H4}$ in a STO-3G basis are plotted at R=1.7\r{A} and show clearly that the RHF cusp observed is a crossing of the two states, giving two-fold degeneracy at 90\textdegree. A NOCI with just these two states recovers a smooth lower (green) and upper (not plotted) state.  The corresponding CCSD states (magenta and cyan) are also plotted and also show a crossing at this point.}
\label{H4_CCSD}
\end{figure}

In our results, the ground RHF state can clearly be seen covering all geometries and giving a poor description of the system energy in the dissociation limit.
For the rectangular geometry, this RHF state breaks down into two symmetry broken UHF states at different geometries.
The lower energy UHF state corresponds to a state with the $\alpha$ electron localised over a pair of hydrogens sharing a short edge of the rectangle and the $\beta$ electron on the opposite side, thus maximising the bonding interaction between the pairs of $\ce{H}$ atoms.
In contrast, the higher energy UHF state corresponds to the state with the $\alpha$ and $\beta$ electrons localised over a pair of hydrogens sharing a long edge of the rectangle.
At bond lengths shorter than Coulson-Fischer points, the hUHF states emerge as predicted to provide a continuation to the normal UHF states.
These show complex coefficients as expected and as seen for the hUHF states of $\ce{H2}$.
For the square geometry, both possible arrangements of the $\alpha$ and $\beta$ electrons become equivalent and thus the states become degenerate.
This causes the single four-fold degenerate hUHF state shown emerging from the RHF state.

\begin{figure*}
	\includegraphics[scale=0.80]{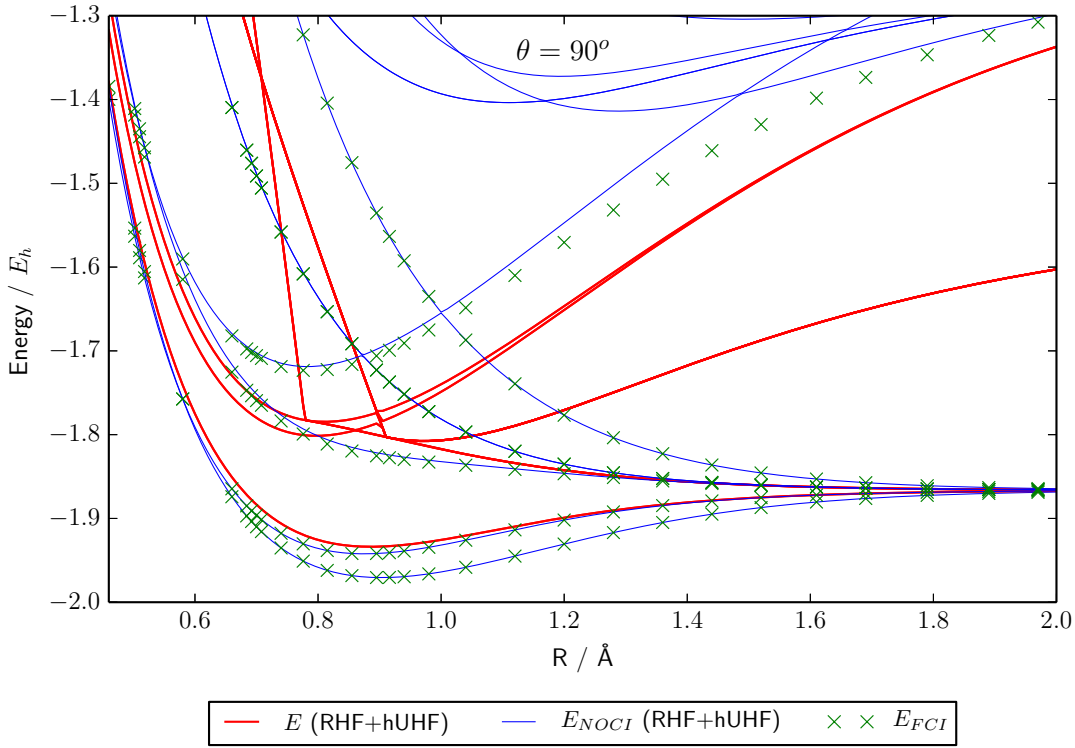}
	\caption{The 18 hUHF states found for square $\ce{H4}$ in a STO-3G basis are plotted in red and used as a basis for generating the NOCI states (blue) for which the ground state corresponds well with the ground FCI state (green). The degeneracy of the hUHF states at R=1.7\r{A} from bottom to top are: 2, 4, 8, 2 and 2.}
	\label{H4_benchmark_90}
\end{figure*}
Using these 5 states (1 $\times$ RHF and 4 $\times$ hUHF) as a basis for non-orthogonal CI produces the states plotted in blue.
Significantly, these states are smooth and continuous across all bond-lengths.
The ground Full CI state is plotted in green and shows good correspondence to the lowest energy NOCI state, which itself is a dramatic improvement on the RHF state.
It is also worth noting that further symmetry breaking of the UHF solutions were identified at larger bond lengths and this may provide a interesting base for further study. 

\section{Holomorphic states of $\ce{H4}$}
We now proceed to examine the SCF solutions of four interacting $\ce{H}$ atoms in the arrangement depicted by Figure \ref{H4_model}.
This model has been used repeatedly to assess the performance of Coupled-Cluster and other methods\cite{HeadGordonBenchmark,ScuseriaH4,H4_DFT} since at large $\theta$ the model resembles well separated $\ce{H2}$ units, where the multireference character of the wavefunction increases, whilst at 90\textdegree\ the four atoms form a square and the solutions become degenerate.
The authors of such studies have repeatedly found what they refer to as a cusp in the RHF energy at 90\textdegree\ which then produces subsequent cusps in the Coupled-Cluster energies generated using this RHF reference.

\begin{figure*}
        \includegraphics[scale=0.42]{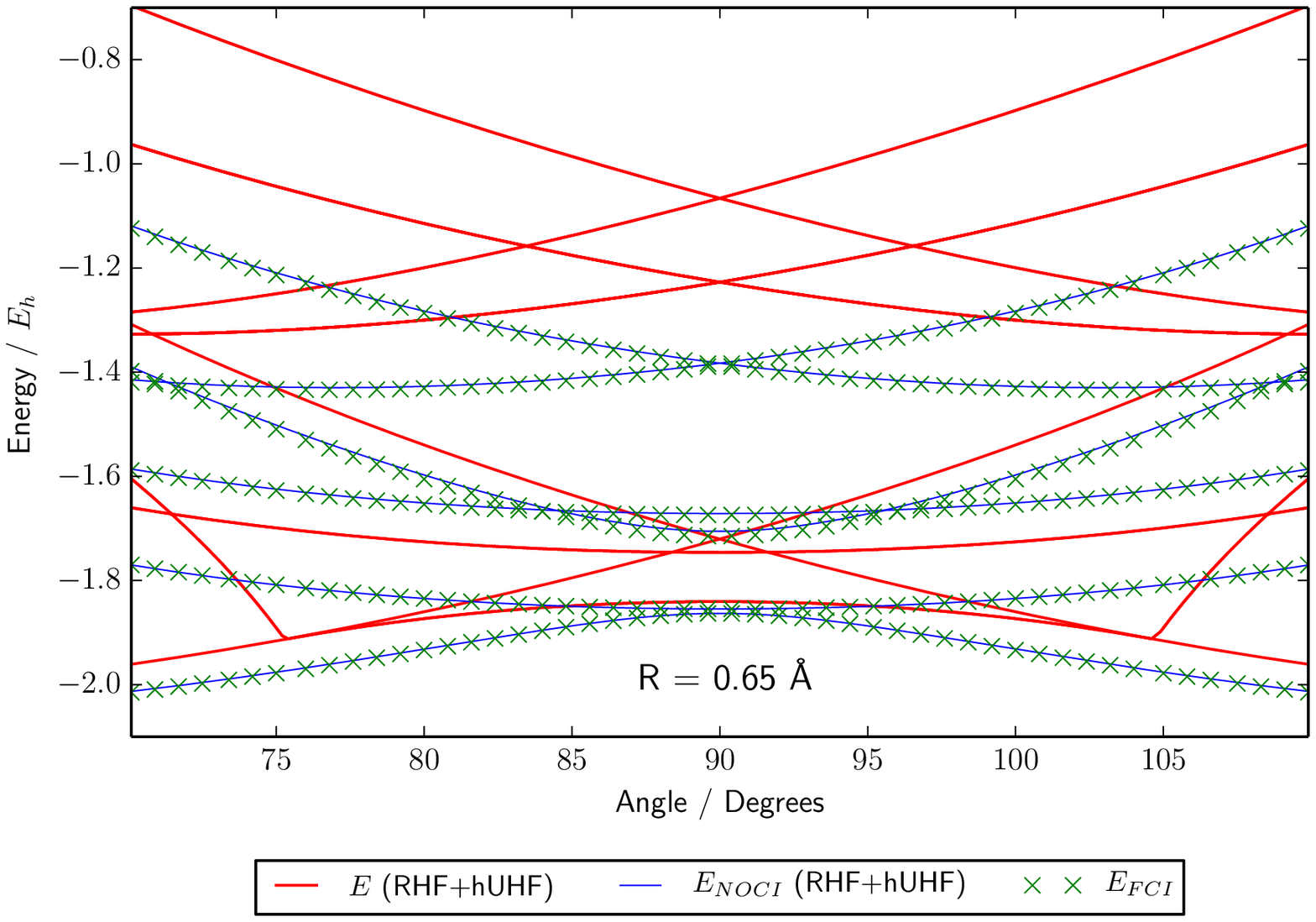}
        \includegraphics[scale=0.42]{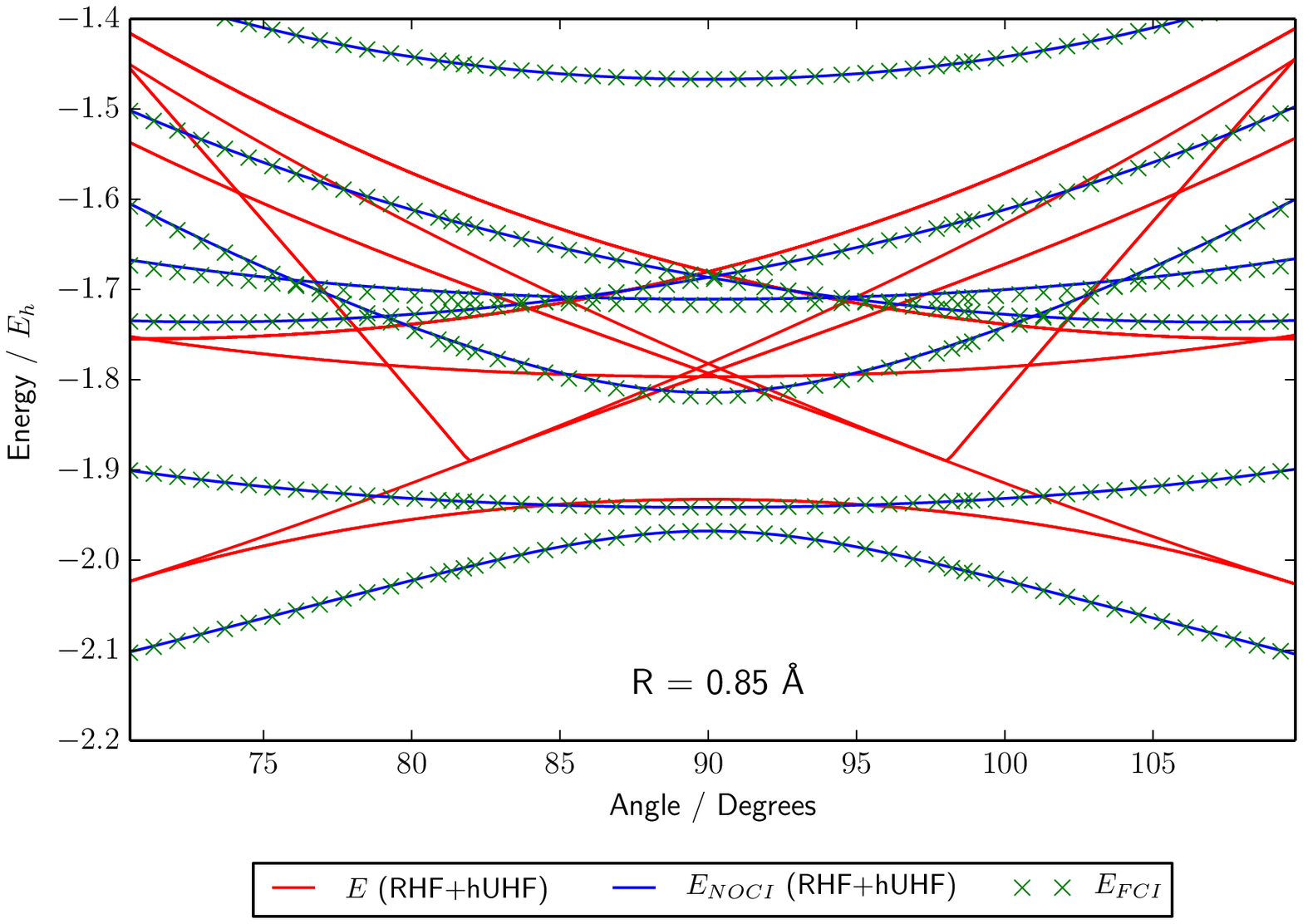}

        \includegraphics[scale=0.45]{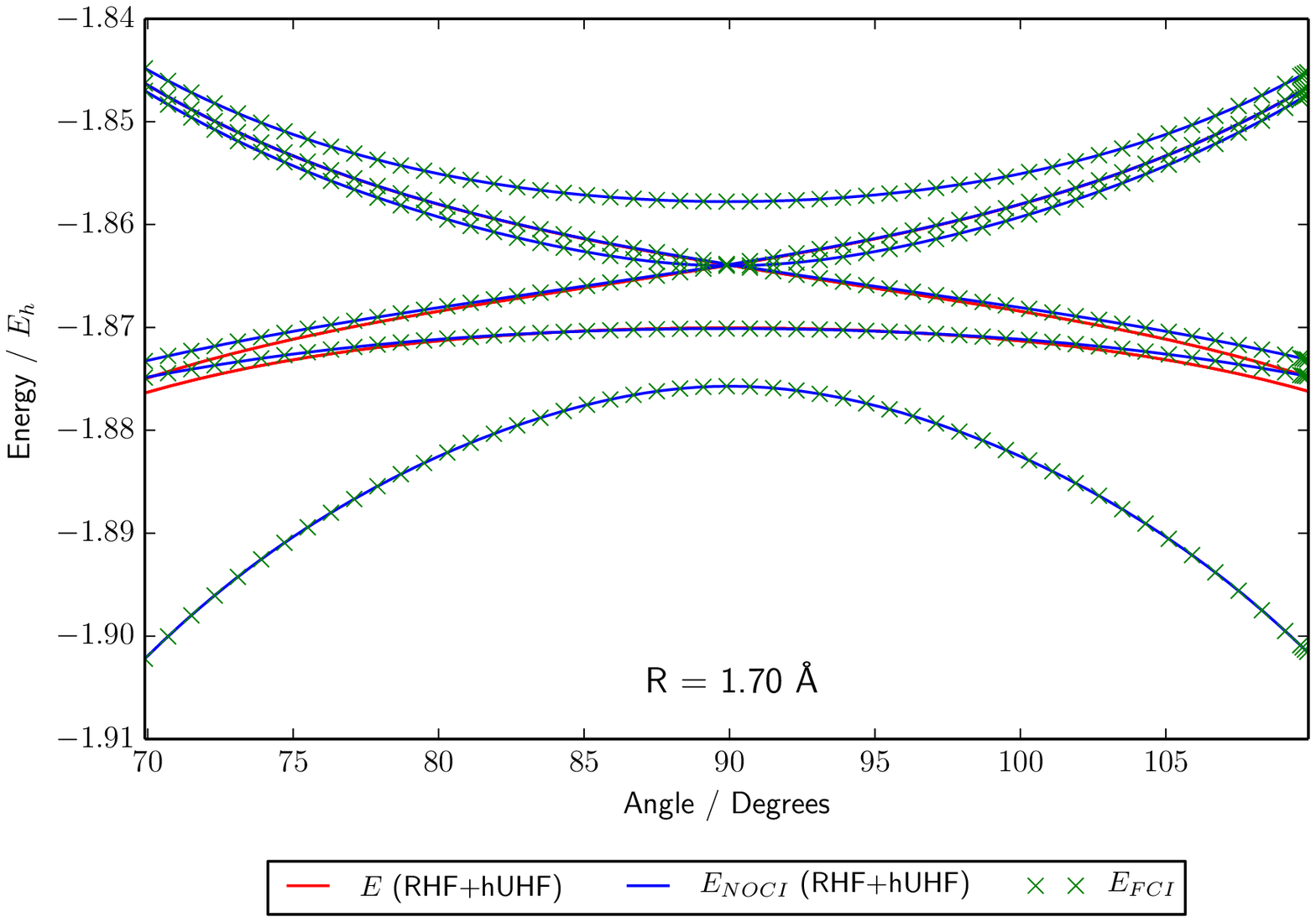}
    	\caption{Cross sections through the $\ce{H4}$ potential energy surface in STO-3G with changing $\theta$ are plotted at R=0.65\r{A}, 0.85\r{A} and 1.70\r{A} and the hUHF states are used to generate NOCI states which show good correspondence to the FCI energies. Significantly, all of the NOCI states plotted are continuous and smooth through 90\textdegree\ and mirror the FCI energies almost exactly.}
	\label{angle_cross_sections}
\end{figure*}

We have replicated the calculations carried out by Scuseria \textit{et al.} \cite{ScuseriaH4} for this system at R=1.70\r{A} but using a STO-3G basis instead of Dunning's DZP basis set, and we plot the two lowest energy RHF states along with their corresponding CCSD energies and the FCI energy in Figure \ref{H4_CCSD}.
In the Scuseria study, only the lowest energy RHF state at each geometry has been used as the reference for the CCSD calculation, ignoring any higher energy solutions.
It appears obvious to us that one cannot ignore these higher energy states since Figure \ref{H4_CCSD} demonstrates that this cusp is by no means an error in the RHF but is a crossing of two solutions, found to be degenerate at 90\textdegree.
Furthermore, it is clear that the subsequent cusp observed in the CCSD calculation is a crossing of two states in an analogous manner to the RHF states. 
Using these two RHF states as a basis for our Non-Orthogonal CI method returns a smooth curve without any cusp.
We note that though the shape models the FCI curve well, the energy of this NOCI solution is significantly too high, indicating that the large contribution due to dynamical correlation is absent.
Using our holomorphic SCF algorithm we identified a further 16 hUHF states of similar or lower energy to these RHF states and these are plotted along the $\theta = 90$\textdegree\ cross-section of the potential energy surface in Figure \ref{H4_benchmark_90}.
Again we observe the emergence of a holomorphic state with complex coefficients and the same degeneracy at the coalescence point of a UHF state with an RHF state, conserving the total number of states.
We apply the NOCI method to the basis set formed from the 18 RHF and hUHF solutions to generate NOCI solutions plotted in blue.
The ground state NOCI state shows good correspondence to the FCI ground state solution, despite the NOCI space only containing 18 determinants compared to the 36 determinants of the Hilbert space.
As a control we have compared to the NOCI formed from the RHF and 17 random complex UHF states (see supplementary Figure 1) where we have found the minimum deviation from the FCI to be 0.1\,$E_h$. 
Further to this, we have also plotted cross-sections through the potential surface between 70 - 110\textdegree\ at R=0.65\r{A}, 0.85\r{A} and 1.70\r{A} using the same hUHF states and these are shown in Figure \ref{angle_cross_sections}. 
These cross sections contain 12, 8 and 0 complex holomorphic states at 90\textdegree\ respectively.
The two lowest NOCI solutions calculated from this hUHF basis are smooth and correspond well to the two lowest FCI states, again showing no cusp at 90\textdegree.
It is also interesting to note that as one moves away from 90\textdegree\ in the R=0.85\r{A} plot, we see the 4 fold degenerate UHF turn holomorphic whilst in the R=0.65\r{A} we see the 2-fold degenerate UHF state also yield a holomorphic state.
Overall, each hUHF solution coalesces with an RHF solution at some point on this potential energy surface.
In future studies we hope to map these coalescence points across the potential energy surface.

\section{Conclusion}
In this study we have demonstrated that using a matrix-driven holomorphic SCF algorithm, it is possible to locate holomorphic-UHF solutions. 
These hUHF solutions form an extension to the conventional UHF solutions where such states coalesce with an RHF state and disappear.
Despite being located by finding the stationary states on a holomorphic, complex-valued potential energy surface, the hUHF states have real-valued holomorphic energy expectation values, $\widetilde{E}$, although we are unsure why this is the case.
Furthermore the hUHF states provide an excellent basis for Non-Orthogonal CI, yielding states which provide an excellent representation of the ground FCI state at a much lower computational cost.
Since the UHF solutions are size extensive, we believe the NOCI states will also be and we hope to verify this in the future.

Using this algorithm, we have been able to demonstrate that the cusps observed in the energy calculations on $\ce{H4}$ actually represents the crossing of two RHF states and the ground FCI state may again be well represented using the hUHF states as a basis for NOCI.
In future studies, we hope to explore the nature of these holomorphic solutions further and look to use the hUHF solutions as a reference for other correlation methods such as Coupled-Cluster. 

AJWT thanks the Royal Society for a University Research Fellowship (UF110161) and HGAB thanks the Royal Society of Chemistry for an Undergraduate Research Bursary.
\bibliography{holoHF}
\end{document}